\documentclass[doublespacing]{elsart}
\usepackage{graphicx}

\begin{document}

\begin{frontmatter}

\title{$S$-wave pairing symmetry in non-centrosymmetric superconductor Re$_3$W}

\author{Jing Yan}, {Lei Shan}, {Qiang Luo}, {Weihua Wang},
\author{Hai-Hu Wen\corauthref{cor}}
\corauth[cor]{Corresponding author.} \ead{hhwen@aphy.iphy.ac.cn}

\address{National Laboratory for Superconductivity,
Institute of Physics and Beijing National Laboratory for Condensed
Matter Physics, Chinese Academy of Sciences, P.~O.~Box 603, Beijing
100080, P.~R.~China}

\begin{abstract}
The alloys of non-centrosymmetric superconductor, Re$_3$W, which
were reported to have an $\alpha$-Mn structure [P. Greenfield and P.
A. Beck, J. Metals, N. Y. \textbf{8}, 265 (1959)] with
$T_\mathrm{c}=9\;$K were prepared by arc melting. The ac
susceptibility and low-temperature specific heat were measured on
these alloys. It is found that there are two superconducting phases
coexisting in the samples with $T_\mathrm{c1}\sim9\;$K and
$T_\mathrm{c2}\sim7\;$K, both of which have a non-centrosymmetric
structure as reported previously. By analyzing the specific heat
data measured in various magnetic fields, we found that the absence
of the inversion symmetry does not lead to the deviation from a
$s$-wave pairing symmetry in Re$_3$W.
\end{abstract}


\end{frontmatter}

 \maketitle

\section{Introduction}
Very recently the scientific community has paid a lot of attention
in understanding the supercondictivity of the non-centrosymmetric
superconductors, since the superconducting properties of such
materials are expected to be unconventional
\cite{CePtSi1,LiPRL1,CePtSi2,Yuan2006,LiPRL2ZhengGQ,Theory,MuG}. In
a lattice with inversion symmetry, the orbital wave function of the
cooper pair has a certain symmetry and the spin paring will be
simply in either the singlet or triplet state. The noncentrosymmetry
in the lattice may bring a complexity to the symmetry of orbital
wave function. This effect with the antisymmetric spin-orbital
coupling gives rise to the broken of the spin degeneracy, thus the
existence of the mixture of spin singlet and triplet may become
possible\cite{LiPRL1,LiPRL2ZhengGQ}. So there might be something
unconventional, such as spin triplet pairing component, existing in
the non-centrosymmetric superconductors. Recently, a spin-triplet
pairing component was demonstrated in Li$_{2}$Pt$_{3}$B both by
penetration depth measurement\cite{Yuan2006} and nuclear magnetic
resonance (NMR)\cite{LiPRL2ZhengGQ}, as was ascribed to the large
atomic number of Pt which enhances the spin-orbit coupling.

Re$_3$W is one of the rhenium and tungsten alloys' family. Up to
now, two superconducting phases of Re$_3$W were reported with
$T_\mathrm{c}\sim9\;$K\cite{HulmJPCS} and
$T_\mathrm{c}\sim\;$7K\cite{Thompson}. Both phases belong to the
$\alpha$-Mn phase (A12, space group I\={4}3m)\cite{XRD}, which has a
non-centrosymmetric structure. Moreover, atomic numbers of Re and W
are 75 and 74, respectively, being close to that of Pt. Therefore,
similar spin-triplet pairing component as that in  Li$_{2}$Pt$_{3}$B
are expected in Re$_3$W. Most recently, it was found that the
superconducting phase of Re$_3$W with $T_\mathrm{c}\sim7\;$K is a
weak-coupling s-wave BCS superconductor by both penetration depth
\cite{Thompson} and Andreev reflection measurements \cite{Huangy}.

In this paper, we report the measurements of the ac susceptibility
and low-temperature specific heat of Re$_3$W alloys. Both the
measurements imply that our samples have two superconducting phases
with critical temperatures near $9\;$K and $7\;$K, respectively, and
the high temperature phase near $9\;$K accounts for nearly 78\%-87\%
in total volume. The specific heat data can be fitted very well by
the simple two-component model, which is based on the isotropic
s-wave BSC theory. Furthermore, a linear relationship is found
between the zero-temperature electronic specific heat coefficient
and the applied magnetic field. These results suggest that the
absence of the inversion symmetry does not result in novel pairing
symmetry in Re$_3$W.

\section{Experiment}

The Re$_3$W alloys are prepared by arc melting the Re and W powders
(purity of 99.9\% for both) with nominal component $3:1$ in a
Ti-gettered argon atmosphere. Normally, the obtained alloy is a
hemisphere in shape with a dimension of $5\;$mm (radius) $\times$
$5\;$mm (height). Some pieces of the alloy had been cut from the
original bulk (e.g. sample~$\sharp1$  and sample~$\sharp2$). The ac
susceptibility of these samples has been measured at zero dc
magnetic field to identify their superconducting phases, whereas,
all of them have two superconducting transitions at about $9\;$K and
$7\;$K, as shown in Fig.~\ref{fig:fig1}. The specific heat was
measured by a Physical Property Measurement System (PPMS, Quantum
Design). The data at a magnetic field were obtained with increasing
temperature after being cooled in field from a temperature well
above $T_\mathrm{c}$, namely, field cooling process.

\section{Results and discussion}

\begin{figure}
\centering
\includegraphics[width=8cm] {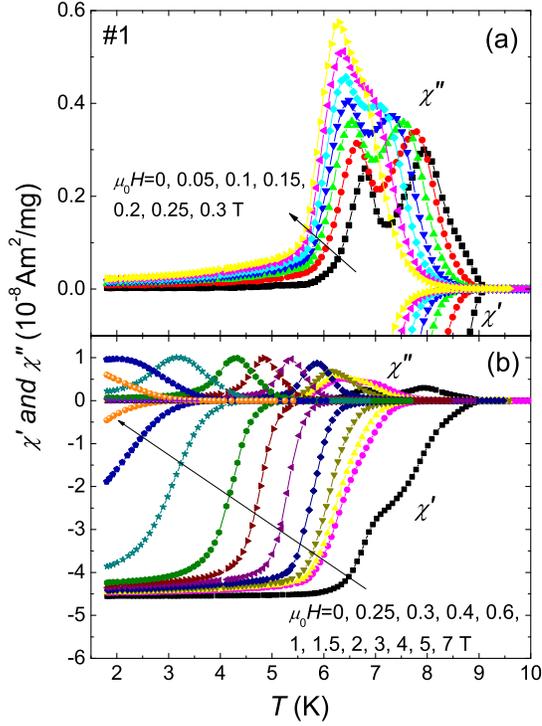}
\caption{(Color online) Temperature dependence of ac susceptibility
on sample~$\sharp1$ under different dc magnetic fields, with ac
field $h=1\;$Oe and frequency $f=333\;$Hz.} \label{fig:fig1}
\end{figure}

The temperature dependence of ac susceptibility ($\chi=\chi' + i
\chi''$) at different dc magnetic fields from $0\;$T to $7\;$T is
shown in Fig.~\ref{fig:fig1}.  One can see that two distinct
superconducting transitions occur at $T_\mathrm{c1}\sim9$ and
$T_\mathrm{c2}\sim 7$ K in $\chi'(T)$ curve at $H=0$
[Fig.~\ref{fig:fig1}(b)], and double peaks in $\chi''(T)$ show up at
the corresponding temperatures. These two phases are consistent with
the previous reports in which they are proofed to be
non-centrosymmetric\cite{HulmJPCS,Thompson}.  The peaks of $\chi''$
shift to lower temperatures as the magnetic field increases, showing
the continuous suppression of superconductivity by the magnetic
field. The low-$T$ peak shifts to lower temperatures more slowly
than the high-$T$ one, indicating distinct behaviors of the upper
critical fields in these two superconducting phases. As $H$
increases to $\sim$ 7 T,  the $\chi(T)$ curves are completely flat,
showing no sign of superconducting transition. Similar results were
obtained on sample $\sharp 2$ and other samples.

\begin{figure}
\centering
\includegraphics[width=8cm] {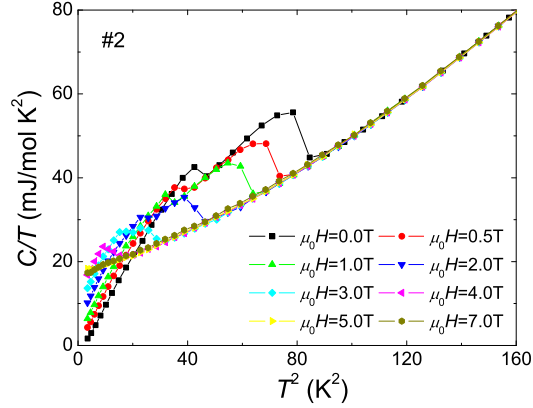}
\caption{(Color online) Specific heat data of sample $\sharp 2$
plotted as $C/T$ versus $T^2$ at various fields.} \label{fig:fig3}
\end{figure}

We thus measured the specific heat of sample $\sharp 2$ and in
Fig.~\ref{fig:fig3} we present the data of $C/T$ versus $T^2$ at
various magnetic fields. On each curve, there are two jumps related
to the superconducting transitions consistent with the measurements
of ac susceptibility. From the zero field data in low temperature
region, one can see that the residual specific heat coefficient
$\gamma_0$ is close to zero, implying the absence of
non-superconducting phase. The superconducting anomaly is suppressed
gradually with increasing magnetic field, and from the curve at
$7\;$T there is no sign of superconductivity above $1.8\;$K,
consistent with the observation in $\chi(T)$ curve.  The low
temperature part of the normal state specific heat at $H=7\;$T in
Fig.~\ref{fig:fig3} is not a straight line, implying that the
specific heat of phonon does not satisfy the Debye's $T^3$ law. We
may need a $T^7$ term to fit the normal state specific heat well:
\begin{equation}
C_\mathrm{n}/T=\gamma_\mathrm{n}+\beta_3T^2+\beta_5T^4+\beta_7T^6.
\label{eq:eq1}
\end{equation}
The first term is the electronic specific heat in the normal state,
and the others are the contributions of the phonons. Fitting the
data of $7\;$T to Eq.~(\ref{eq:eq1}), the coefficients
$\gamma_\mathrm{n}=17\pm0.1\;$mJ/mol$\;$K$^2$,
$\beta_3=0.185\pm0.001\;$mJ/mol$\;$K$^4$,
$\beta_5=(1.63\pm0.01)\times10^{-3}\;$mJ/mol$\;$K$^6$, and
$\beta_7=(-2.087\pm0.005)\times10^{-6}\;$mJ/mol$\;$K$^8$ are
determined. From the relation:
\begin{equation}
\beta_3=\frac{12\pi^4}{5}\frac{N_\mathrm{A}k_\mathrm{B}Z}{\Theta^3_\mathrm{D}},
\label{eq:eq2}
\end{equation}
where $N_\mathrm{A}=6.02\times10^{23}$ is the Avogadro constant, and
$Z=4$ the number of atoms in one unit cell, we obtained the Debye
temperature of our alloys $\Theta_\mathrm{D}=347.9\;$K. These
coefficients and Debye temperature are all very close to the results
of other works on Re-W
alloys\cite{MullerHPA,RevModPhys,SHPhysRev,MullerSH,MullerLowTemp,WChu,Stewart1978}.

\begin{figure}
\centering
\includegraphics[width=8cm] {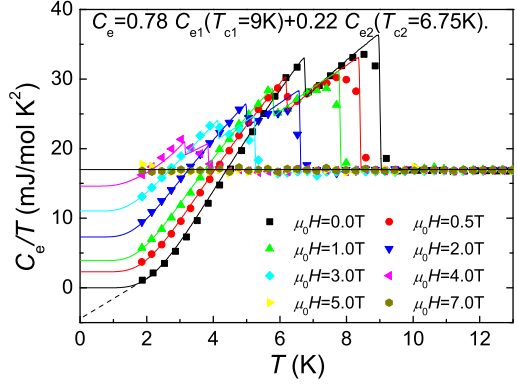}
\caption{(Color online) Specific heat of electrons plotted as
$C_\mathrm{e}/T$ versus T. The solid lines are the calculating
results which separate the electronic specific heat into two
components with different $T_\mathrm{c}$ by using specific heat
formula based on the BCS theory.} \label{fig:fig4}
\end{figure}

By subtracting the phonon contribution, the electronic specific heat
$C_\mathrm{e}$ is obtained, which is shown in Fig.~\ref{fig:fig4} as
$C_\mathrm{e}/T$ versus $T$. Before a quantitative analysis, the low
temperature specific heat at low fields has presented a strong
evidence that Re$_3$W has a nodeless gap function. For a nodal
superconductor (expected by the strong mixing of spin-singlet and
spin-triplet pairing components in a heavily non-centrosymmetric
superconductor such as Li$_2$Pt$_3$B), the low temperature $C/T$ vs.
$T$ relation should be a power law like. However, as denoted by the
dashed lines in Fig.~\ref{fig:fig4}, if a linear relationship is
assumed, the specific heat at zero field would be negative when the
temperature approaches to zero. In the following section, by using a
quantitative analysis, we will demonstrate that both phases of
Re$_3$W have an isotropic gap function, which is in good agreement
with the expectation of an $s$-wave superconductor.

\begin{figure}
\centering
\includegraphics[width=8cm] {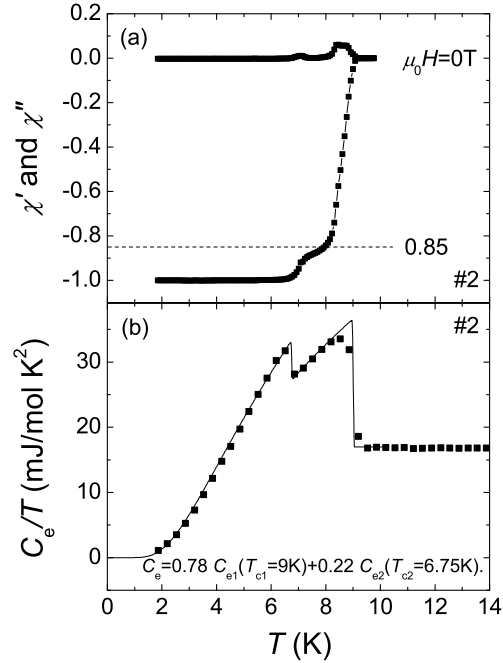}
\caption{(a) shows the ac susceptibility of sample $\sharp2$ on
which the specific heat have been measured. (b) shows the zero field
specific heat data, and the black line is the calculating result
based on the BCS theory.} \label{fig:fig5}
\end{figure}

Figure~\ref{fig:fig5} shows the ac susceptibility and specific heat
data at zero dc field measured on the same sample($\sharp2$). The ac
susceptibility data have been normalized. The high temperature phase
occupies nearly $85\pm 1$\% in the whole superconducting volume. In
order to fit the zero field electronic specific heat, we attempt to
use the formula derived from thermodynamic relations based on the
BCS theory\cite{Tinkham}
\begin{equation}
C_\mathrm{es}=\frac{4N(0)}{k_\mathrm{B}T^{2}}\int_{0}^{\infty}
\frac{e^{\zeta/k_\mathrm{B}T}}{(1+e^{\zeta/k_\mathrm{B}T})^{2}}
(\varepsilon^{2}+\Delta^{2}(T)-\frac{T}{2}\frac{\mathrm{d}\Delta^{2}(T)}
{\mathrm{d}T})\,\mathrm{d}\varepsilon, \label{eq:eq3}
\end{equation}
where $\zeta=\sqrt{\varepsilon^2+\Delta^2(T)}$, and $\Delta(T)$ is
an isotropic $s$-wave gap which depends on temperature in the same
way as expected by BCS theory. Since there are two coexistent phases
in our samples, we use two separate terms of $C_H$ and $C_L$ to take
into account the contributions of the high $T_\mathrm{c}$ and low
$T_\mathrm{c}$ phases, respectively. Thus the total specific heat
can be expressed as follows:
\begin{equation}
C_\mathrm{e}=\omega_HC_H+\omega_LC_L, (\omega_H+\omega_L=1),
\label{eq:eq5}
\end{equation}
in which $\omega_H$ and $\omega_L$ indicate the weight of the
contributions for the two phases. According to Eq.~(\ref{eq:eq3})
and Eq.~(\ref{eq:eq5}) we can nicely simulate the experimental data
very well as presented in Fig.~\ref{fig:fig5}(b) by a solid line.
The parameters for the best fit are $\Delta_{0H}=1.4\;$meV,
$\omega_H=0.78$ for $T_{\mathrm{c}H}=9\;$K and
$\Delta_{0L}=1.1\;$meV, $\omega_L=0.22$ for
$T_{\mathrm{c}L}=6.75\;$K and $\Delta_0$ is the gap value at zero
temperature. Interestingly, $\omega_H=0.78$ found here is very close
to the relative weight 85\% of the high temperature phase which was
obtained from the ac susceptibility data in Fig.~\ref{fig:fig5}(a).
Furthermore, $\Delta_{0L}\sim1.1\;$meV is in good agreement with
that from the penetration depth and Andreev reflection
experiments\cite{Thompson,Huangy}. These results give a strong
evidence that there is no novel pairing symmetry in our alloys.

To get further evidence for this argument, we did similar
calculations for the specific heat in the mixed state using the same
weights of the two phases obtained from the zero field calculation.
In the mixed state, there are two different regions, namely the core
region and the outside core region. Therefore we adopted a simple
two-component model\cite{Twochannel,LShan} which separates the
electronic specific heat into two components. The electronic
specific heat is thus written as
\begin{equation}
C_\mathrm{e}=\alpha\frac{H}{H_{\mathrm{c}2}(0)}\gamma_\mathrm{n}T+(1-\alpha\frac{H}{H_{\mathrm{c}2}(0)})C_\mathrm{es}.
\label{eq:eq4}
\end{equation}
Here $\alpha$ is an adjustable parameter. The first part on the
right hand side is the quasi-particle density of states (DOS) coming
from the normal vortex core regions, and the second part comes from
the superconducting regions outside the cores. The results of the
quantitative calculations are plotted as solid lines in
Fig.~\ref{fig:fig4}, and they are in good agreement with the
experimental data for all magnetic fields.

\begin{figure}
\centering
\includegraphics[width=8cm] {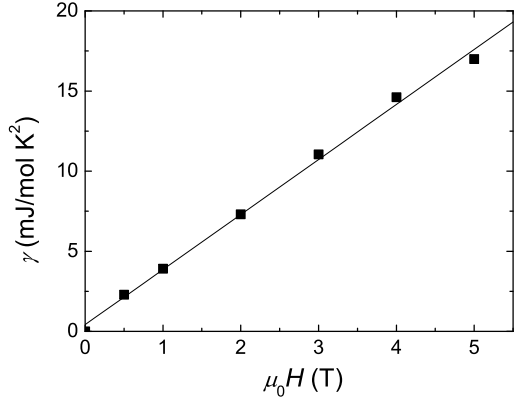}
\caption{The electronic specific coefficient $\gamma(H)$ at zero
temperature obtained from the calculation based on BCS theory.}
\label{fig:fig7}
\end{figure}

In the superconducting state, $C_\mathrm{e}=\gamma T$, where
$\gamma$ is the electronic specific heat coefficient that is
dependent on temperature and field. According to Eq.~(\ref{eq:eq4}),
the zero temperature electronic specific heat coefficient
$\gamma(H)$ is equal to $\alpha
H/H_\mathrm{c2}(0)\gamma_\mathrm{n}$, which is shown in
Fig.~\ref{fig:fig7} as solid squares, and the solid line is a linear
fit to the data. The obvious linear relationship of $\gamma$ vs. $H$
presents further evidence that Re$_3$W is a conventional
superconductor in which $\gamma(H)$ is proportional to the number of
vortex cores and hence to the applied field. For a nodal
superconductor with novel pairing symmetry, on the other hand, a
nonlinear $\gamma(H)$ relation should be expected, which is
obviously not the case in our present samples\cite{HfV2}.

\section{Conclusion}
In summary, we have synthesized Re$_3$W alloys by arc melting. From
the measurements of ac susceptibility and specific heat on the
alloys two distinct superconducting phases were found. Both the
qualitative and quantitative analysis were done on the specific heat
data in zero field and the mixed state. We found that the simple
two-component model based on the BCS theory with an isotropic
$s$-wave gap can fit our experimental data very well, and we
obtained a linear $\gamma(H)$ relationship. All these results
indicate that the absence of the inversion symmetry does not result
in any novel pairing symmetry in Re$_3$W for both
$T_\mathrm{c}\sim7\;$K and $T_\mathrm{c}\sim9\;$K phases.

\section*{Acknowledgments}
This work was supported by the National Science Foundation of China,
the Ministry of Science and Technology of China (973 Project: No.
2006CB601000, No. 2006CB921802, No. 2006CB921300), the Knowledge
Innovation Project of Chinese Academy of Sciences (ITSNEM).

\end{document}